\documentclass[submission,copyright]{eptcs}
\providecommand{\event}{SSV 2012}
\usepackage{breakurl}
\usepackage[pdftex]{graphicx}

\newif \ifDraft         \Draftfalse
\newif\ifFinal \Finaltrue

\setkeys{Gin}{keepaspectratio=true,clip=true,draft=false,width=\linewidth}
\graphicspath{{./imgs/}}

\ifDraft
  \usepackage{draftcopy}
  \newcommand{\Comment}[1]{\textbf{\textsl{#1}}}
  \newcommand{\FIXME}[1]{\textbf{\textsl{FIXME: #1}}}
\else
  \newcommand{\Comment}[1]{\relax}
  \newcommand{\FIXME}[1]{\relax}
\fi

\usepackage[T1]{fontenc}
\usepackage[utf8]{inputenc}

\usepackage{cite}

\usepackage{amsmath}
\usepackage{amssymb}
\usepackage{stmaryrd}
\usepackage{tensor}
\usepackage{mathpartir}

\newcommand{\Abort}{\textrm{Abort}}
\newcommand{\Skip}{\textrm{Skip}}
\newcommand{\Apply}{\textrm{Apply}}
\newcommand{\Embed}{\textrm{Embed}}
\newcommand{\Seq}{\mathbin{;;}}
\newcommand{\Lmbd}[2]{\lambda #1.\ #2}
\newcommand{\PC}[1]{\tensor[_{#1}]{\oplus}{}}
\newcommand{\DC}{\sqcap}
\newcommand{\Mu}[2]{\mu #1.\ #2}
\newcommand{\EB}[1]{\text{«} #1 \text{»}}
\newcommand{\Wp}{\mathop{\text{wp}}}
\newcommand{\Wlp}{\mathop{\text{wlp}}}
\newcommand{\If}{\textrm{if}}
\newcommand{\Then}{\textrm{then}}
\newcommand{\Else}{\textrm{else}}
\newcommand{\entails}{\vdash}
\newcommand{\pentails}{\Vdash}
\newcommand{\valid}[3]{\lbrace #1 \rbrace\ #2\ \lbrace #3 \rbrace}
\newcommand{\refinedby}{\sqsubseteq}

\usepackage{isabelle,isabellesym}
\isabellestyle{rm}

\begin{document}

\renewcommand{\sectionautorefname}{Section}
\renewcommand{\subsectionautorefname}{Section}
\renewcommand{\subsubsectionautorefname}{Section}
\renewcommand{\appendixautorefname}{Appendix}
\renewcommand{\Hfootnoteautorefname}{Footnote}
\newcommand{\Htextbf}[1]{\textbf{\hyperpage{#1}}}

\title{Verifying Probabilistic Correctness\\in Isabelle with {pGCL}}

\author{
    David Cock
    \institute{NICTA\\ Sydney, Australia\thanks{\scriptsize NICTA is funded 
     by the Australian Government as represented by the Department of 
     Broadband, Communications and the Digital Economy and the Australian Research
     Council through the ICT Centre of Excellence program.}}
    \institute{School of Computer Science and Engineering\\
               University of New South Wales}
    \email{David.Cock@nicta.com.au}
}

\def\titlerunning{Verifying Probabilistic Correctness in Isabelle with {pGCL}}
\def\authorrunning{David Cock}

\maketitle

\urlstyle{sf}
\thispagestyle{empty}
\begin{abstract}
This paper presents a formalisation of pGCL in Isabelle/HOL.  Using a shallow
embedding, we demonstrate close integration with existing automation support.
We demonstrate the facility with which the model can be extended to
incorporate existing results, including those of the L4.verified project.  We
motivate the applicability of the formalism to the mechanical verification of
probabilistic security properties, including the effectiveness of side-channel
countermeasures in real systems.
\end{abstract}

\section{Introduction}\label{s:intro}

We present a new formalisation of the pGCL programming logic within
Isabelle/HOL\cite{Nipkow_PW:Isabelle}.  The motivation for this work is the
desire for formal, mechanised verification of probabilistic security
properties (in particular, bounds on side-channel vulnerability) for real
systems.  We build on existing theoretical work on pGCL\cite{McIver_M_04}, and
present a complementary approach to the existing formalisation in
HOL4\cite{Hurd_05}.

Our contribution is a shallow embedding of pGCL within Isabelle/HOL, using the
unmodified real-number type for probabilities, leading to excellent support
for proof automation and easy extensibility to tackle novel problems, and to
integrate with existing results (including those of
L4.verified\cite{Klein_EHACDEEKNSTW_09}).  We demonstrate these by example.

The structure of the paper is as follows: We first motivate the need for
probabilistic reasoning in systems (\autoref{s:why}), and in particular the
need for probabilistic (\autoref{s:sec_prop}) and refinement-sound
(\autoref{s:refine}) security properties.  We then present our pGCL theory
package (\autoref{s:what}), giving a cursory introduction to its syntax and
underlying model (\autoref{s:etmodel}), and demonstrate the use of the package
(\autoref{s:using}), giving examples of 3 styles of proof.  We next provide
more detail on the underlying implementation (\autoref{s:how}), touching on
the specifics of the shallow embedding (\autoref{s:wp_impl}), extending it
into novel contexts (\autoref{s:shallow}), the lattice structure of the
semantic models that underlie recursion (\autoref{s:lattice}), and finally the
implementation of the VCG (\autoref{s:vcg}).  Surveying related work
(\autoref{s:related}), we conclude by describing the ongoing efforts in
applying the tool (\autoref{s:ongoing}).

\section{The Case for Probabilistic Correctness}\label{s:why}

Recent work\cite{Klein_EHACDEEKNSTW_09} has demonstrated the practicality of
verifying the functional correctness of real systems.  Functional correctness,
however, covers only properties that are {\em guaranteed} to hold.  This
excludes classes of properties relevant in practice, including certain
execution-time and security guarantees.  By allowing security properties that
only hold with some probability, we can give a more nuanced classification of
systems than that arising from a functional property such as
noninterference\cite{Goguen_Meseguer_82}.

\subsection{Probabilistic Behaviour in Systems}

In formal specification, it is convenient to treat a system as fully
predictable, and make concrete truth claims about its behaviour.  Once
implemented, while the system may be in principle predictable, the size and
complexity (and often under-specification) of the state space makes a full
description impractical.

The usual approach in this case is to retreat to a probabilistic model of the
system, informed by benchmarking.  That this is true is apparent on reading
the evaluation section of a systems paper:  While the precise performance of a
system is theoretically predictable and could thus be calculated without
empiricism,  what we see are benchmarks and histograms.  The tools of the
evaluator are experiments and statistics!  This is a testament to the immense
difficulty of precisely predicting performance.  Once real world phenomena
intrude, as in networking, exact prediction becomes impossible, even in
theory.

\begin{figure}
    \begin{center}
        \includegraphics{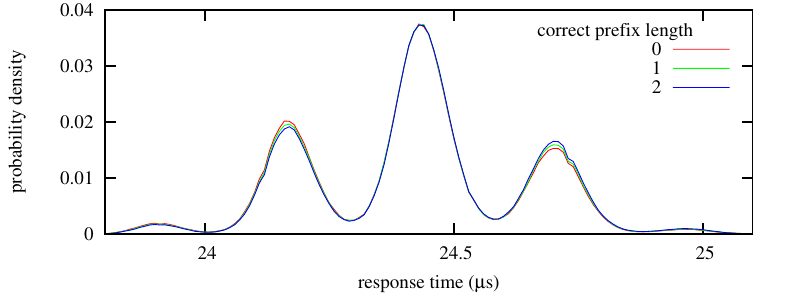}
    \end{center}
    \caption{\label{f:strcmp}String compare ({\tt strcmp}) execution time
             distribution}
\end{figure}

Moreover, there exist properties which can only be treated with statistical
methods.  If such properties are critical for correctness, then a proof will,
of necessity, involve probabilistic reasoning.  Execution time is an example.
\autoref{f:strcmp} is representative, giving the response time distribution
for the C {\tt strcmp} routine for a fixed secret and a varying
guess\footnote{This figure first appeared in a previous work\cite{Cock_11},
where it is discussed in more depth.}.  Note that within the measurement
precision ($10ns$) the distribution is essentially continuous, and depends on
a sensitive datum: the longest common prefix between the secret and the guess.
If we are concerned with security, this is a correctness-critical property
that is unavoidably probabilistic.  This is a simple side-channel, vulnerable
to a guessing attack.

\subsection{Security Properties}\label{s:sec_prop}

As stated, our motivation  is the verification of probabilistic security
properties, given a concrete attack model.  Building on the above example,
imagine that an adversary is supplying guesses in decreasing order of
probability, and updating this order based on observed response time.  A
possible security property then, is that the adversary does not guess the true
secret in fewer than, say, $n$ attempts.

For a functional security property, we would now calculate the set of initial
states that guarantee that our property will hold in the final state: its {\em
weakest precondition} ($\Wp$).  Security is assured by showing that the
initial state lies in this set.

In a probabilistic system, however, it could easily be the case that from {\em
any} initial state, there is a nonzero probability that the final state is
insecure, even if the overall chance of this occurring is small.  This would be
the case if the attacker guessed completely at random.  The best we could then
hope for is to find the {\em probability} that the final state is secure.

What we need is a probabilistic analogue of the weakest precondition.
Consider the functional weakest precondition as a $\{0,1\}$-valued function
(identifying it with its selector).  We then ask whether we can define a
$[0,1]$-valued function, that instead of answering ``secure'' or ``not
secure'' for a state, answers ``secure with probability $p$''.  We could then
show that the system remains secure with probability $\ge p$ by showing that
the initial state lies in the set for which $\Wp \ge p$.  In the probabilistic
programming logic pGCL of McIver \& Morgan\cite{McIver_M_04} we find just
such an analogue.

\subsection{Refinement and Security}
\label{s:refine}

The standard problem in applying refinement to security is in treating
nondeterminism. Writing $a\Seq b$ for the sequential composition of programs
$a$ and $b$, and $c\DC d$ for a demonically nondeterministic choice between
$c$ and $d$, consider the following fragment:
$$h := \text{secret} \Seq (l := 0 \DC l := 1)\quad\text{where}\quad
\text{secret}\in\{0,1\}\ \text{.}$$
Let $h$ (high) be hidden, and $l$ (low) be globally visible. We wish to
formalise the intuitive security property: `the value in $l$ doesn't reveal
the value in $h$'.  Writing $a \refinedby b$ for `$b$ refines $a$', taken to
mean that every trace of $b$ is also a trace of $a$, the following is a valid
refinement:
$$h := \text{secret} \Seq l := h\ \text{,}$$
which clearly violates the security property.  Any attempt to verify a
property such as $\forall s_\text{final}.\ l \neq h$ for this program fails
due to such insecure refinements.\footnote{With a state space of two, of
course, such an anti-correlation would be just as insecure as $l=h$.  In a
large space however, knowing $l\neq h$ leaks very little information.  We are
not establishing that $l$ and $h$ are uncorrelated, rather that in a given
trace they did not happen to coincide.  When we consider guessing attacks,
this will be precisely the formulation we need: Whether or not the attacker
simply got lucky is irrelevant, as the system is still compromised.}

For a property $Q$ to survive refinement, writing $\entails$ for predicate
entailment, we need that
$$
\inferrule*[lab=,sep=1em]
    {a \refinedby b}
    {\Wp\ a\ Q \entails \Wp\ b\ Q}
\qquad\text{whence by transitivity,}\qquad
\inferrule*[lab=,sep=1em]
    {a \refinedby b \\ P \entails \Wp\ a\ Q}
    {P \entails \Wp\ b\ Q} \ \text{.}
$$

Under an appropriate model (e.g.~$\Wp\ (a \DC b)\ P\ s = \Wp\ a\ P\ s \land
\Wp\ b\ P\ s$), these relations hold for the above program and security
predicate, but show the hopelessness of the case as:
$$\forall s.\
    \lnot\Wp\ (h := \text{secret} \Seq (l := 0 \DC l := 1))\ (l\neq h)\ s
\ \text{.}$$

In other words the weakest precondition of $l\neq h$, considered as a set, is
empty, and thus the original specification cannot satisfy the security
property.  Pessimistically, on any trace the program might set $l=h$.  It is
critical to establish that a desired security property is preserved by
refinement, in order that such insecure specifications are exposed.

Our formalism of choice is pGCL, where we have a novel notion of entailment.
We write\footnote{We differ in syntax, as the symbol $\Rrightarrow$ is not
readily available within Isabelle} $P \pentails Q$ for comparison defined
pointwise i.e.~$\forall s.\ P\ s \le Q\ s$.  If $P \pentails Q$, then $P$ has
a lower value in every state than does $Q$.  Write\footnote{Again we differ,
as the established syntax, $[\cdot]$, clashes with lists.} $\EB{P}$ for the
boolean predicate $P$ as real-valued function (an {\em expectation}):
$$
\EB{P}\ s = \If\ P\ s\ \Then\ 1\ \Else\ 0
\quad\text{noting that}\quad
P \entails Q \leftrightarrow \EB{P} \pentails \EB{Q}
$$

Considering again the example of \autoref{s:sec_prop}, we model a guessing
attack on $h$ as
\begin{equation}\label{e:guessing_attack}
(h:=0 \DC h:= 1) \Seq (l:=0 \PC{1/2} l:=1)\ \text{,}
\end{equation}
where the secret ($h$) is chosen nondeterministically, and the guess ($l$)
randomly ($a \PC{p} b$ denotes probabilistic choice between programs $a$ and
$b$, with probability $p$ for $a$).

We have structural refinement rules, for example
$a \DC b \refinedby a$, and the following relations:
$$
\inferrule*[lab=wp\_refines,sep=1em]
    {a \refinedby b}
    {\Wp\ a\ Q \pentails \Wp\ b\ Q}
\quad\text{whence}\quad
\inferrule*[lab=,sep=1em]
    {a \refinedby b \\ P \pentails \Wp\ a\ Q}
    {P \pentails \Wp\ b\ Q}\ \text{,}
$$
which are the probabilistic equivalents of our refinement conditions. We have,
therefore, that $h:=0 \Seq (l:=0 \PC{1/2} l:=1)$ is a refinement of
\autoref{e:guessing_attack}, but importantly, $(h:=0 \DC h:=1) \Seq l:=h$ is
not.  In contrast to demonic choice, probabilistic choice cannot be refined
away.

A refinement in pGCL establishes a predicate with {\em higher} probability
than does its specification.  Thus if we arrange our predicate on final states
as `the system is secure', and we are content with establishing the minimum
probability with which it is established, refinement {\em by definition} can
only increase this probability.  Our guessing-game model is thus
refinement-sound in pGCL.

\section{The {pGCL} Theory Package}\label{s:what}

To automate such reasoning, we present a shallow embedding of pGCL in
Isabelle/HOL.  With a few noted exceptions, we preserve the syntax of McIver
\& Morgan.  The advantage of our approach is the ease with which we can apply
the existing machinery of Isabelle/HOL to the underlying arithmetic.  The
disadvantage is that we cannot appeal to the axioms of a stricter type, in
particular Isabelle's fixed-point lemmas, which reside in the class of
complete lattices.  This is dealt with in detail in \autoref{s:how}.

The language models imperative computation, extended with both
nondeterministic and probabilistic choice.  The non-probabilistic component
corresponds to Dijkstra's guarded command language, GCL\cite{Dijkstra_75}.
Nondeterminism (by default) is demonic, with respect to the postcondition of a
program:  A demonic choice minimises the likelihood of establishing the
postcondition.

Sequential composition ($\Seq$), demonic choice ($\DC$) and probabilistic
choice ($\PC{p}$) were introduced in \autoref{s:refine}.  The remainder
consists of: $\Abort$ and $\Skip$, representing failure and a null operation,
respectively; $\Apply$, which embeds a state transformer; and the recursive
primitive, $\mu$.

\subsection{The Expectation-Transformer Model}
\label{s:etmodel}

The intuitive interpretation of a program, its operational semantics, is as a
probabilistic automaton.  From a given state, the program chooses the next
randomly, with a probability depending on the state.  The program is thus a
{\em forward} transformer, taking a distribution on initial states to one on
final states.

For mechanisation, we are more interested in the axiomatic interpretation of a
program: as a {\em reverse} transformer.  Here, the program maps a function on
the final state to one on the initial state: a `real-valued predicate'.  These
generalised predicates are the {\em expectations} of \autoref{s:refine}, and
are the bounded, nonnegative functions from the state space $S$, to
$\mathbb{R}$:
\begin{align*}
\text{bounded\_by}\ b\ P &\equiv \forall s.\ P\ s \le b &
\text{bounded}\ P &\equiv \exists b.\ \text{bounded\_by}\ b\ P \\
\text{nneg}\ P &\equiv \forall s.\ 0 \le P\ s &
\text{sound}\ P &\equiv \text{bounded}\ P \land \text{nneg}\ P
\end{align*}
The {\em strict} interpretation of non-recursive constructions is given in
\autoref{f:rules}.  The {\em liberal} interpretation differs only for $\Abort$
and $\mu$, and is explained in \autoref{s:wp_impl}. That the forward and
reverse interpretations are equivalent is established\cite{McIver_M_04},
although we have not mechanised the proof.

To see that this model produces the probabilistic weakest precondition
demanded in \autoref{s:why}, note that the expectation $\EB{R} : S \rightarrow
\mathbb{R}$, applied to final states, gives the probability that the predicate
$R$ holds.  This interpretation is preserved under transformation: $\Wp\
\text{prog}\ \EB{R}\ s_\text{initial}$ is the probability that $R$ will hold
in the final state, if prog executes from $s_\text{initial}$.  Equivalently,
it is the current expected value of the predicate on the final state:
$\sum_{s_\text{final}} P(s_\text{final}|s_\text{initial}) \times \EB{R}\
s_\text{final}$, hence the term expectation.

A program is represented as a function from expectation to expectation:
$(S \rightarrow \mathbb{R}) \rightarrow S \rightarrow \mathbb{R}$.
The function maps post-expectation to weakest pre-expectation.  For a {\em
standard} post-expectation, the embedding of a predicate (e.g.~$\EB{P}$), this
is the greatest lower bound on the likelihood of it holding, finally.

For our inference rules to apply, transformers must satisfy several {\em
healthiness} properties, which are slightly weaker than the standard
versions\cite{McIver_01}.  We combine the treatment of strict transformers
(weakest precondition, giving total correctness) and liberal transformers
(weakest {\em liberal} precondition, giving partial correctness) by working in
the union of their domains.  This basic healthiness is defined as the
combination of {\em feasibility}, {\em monotonicity} and {\em weak scaling}:
\begin{gather*}
\textrm{feasible}\ t \equiv
    \forall P\,b.\ \textrm{bounded\_by}\ b\ P \land \textrm{nneg}\ P
                \rightarrow \textrm{bounded\_by}\ b\ (t\ P) \land
                \textrm{nneg}\ (t\ P) \\
\textrm{mono\_trans}\ t \equiv
    \forall P\,Q.\ (\textrm{sound}\ P \land \textrm{sound}\ Q
                    \land P \pentails Q) \longrightarrow
                   (t\ P) \pentails (t\ Q) \\
\textrm{scaling}\ t \equiv
    \forall P\,c\,s.\ (\textrm{sound}\ P \land 0 < c) \longrightarrow
                  c \times t\ P\ s = t\ (\lambda s.\ c \times P\ s)\ s
\end{gather*}
Stronger results are established on-the-fly by appealing to one of several
supplied rule-sets.

A well-defined program has healthy strict and liberal interpretations, related
appropriately:
$$
\textrm{well\_def}\ a \equiv \textrm{healthy}\ (\Wp\ a)
                      \land  \textrm{healthy}\ (\Wlp\ a)
                      \land  (\forall P.\ \textrm{sound}\ P \longrightarrow
                                          \Wp\ a\ P \pentails \Wlp\ a\ P)
$$

\begin{figure}
\begin{center}
\begin{tabular}{r@{$\quad=\quad$}lr@{$\quad=\quad$}l}
$\Wp\ \Abort\ R$ & $\Lmbd{s}{0}$ &
$\Wp\ (a \Seq b)\ R$ & $\Wp\ a\ (\Wp\ b\ R)$ \\
$\Wp\ \Skip\ R$ & $R$ &
$\Wp\ (a \PC{p} b)\ R$ & $\Lmbd{s}{p\ s \times \Wp\ a\ R\ s +
                                   (1 - p\ s) \times \Wp\ b\ R\ s}$ \\
$\Wp\ (\Apply\ f)\ R$ & $\Lmbd{s}{R\ (f\ s)}$ &
$\Wp\ (a \DC b) R$ & $\Lmbd{s}{\min\ (\Wp\ a\ R\ s)\ (\Wp\ b\ R\ s)}$
\end{tabular}
\end{center}
\caption{\label{f:rules}Structural $wp$ rules.}
\end{figure}

\subsection{Reasoning with pGCL}
\label{s:using}
The rules in \autoref{f:rules} evaluate the weakest pre-expectation of
non-recursive program fragments structurally.  If the exponential growth of
the term is not problematic, the simplifier can calculate it exactly.  We also
show two other approaches supported by the package: Modular reasoning by
structural decomposition, and the verification condition generator (VCG).
Examples are given in Isabelle proof script.

\begin{figure}
\begin{align*}
\text{hide} &\equiv \Apply\ \big(\Lmbd{s} s\llparenthesis
                    \text{P} \mathrel{:=} 1\rrparenthesis\big)
                \DC \Apply\ \big(\Lmbd{s} s\llparenthesis
                    \text{P} \mathrel{:=} 2\rrparenthesis\big)
                \DC \Apply\ \big(\Lmbd{s} s\llparenthesis
                    \text{P} \mathrel{:=} 3\rrparenthesis\big)\\
\text{guess} &\equiv       \Apply\ \big(\Lmbd{s} s\llparenthesis
                           \text{G} \mathrel{:=} 1\rrparenthesis\big)        
                \PC{{\Lmbd{s}{1/3}}} 
                           \Big(\Apply\ \big(\Lmbd{s} s\llparenthesis
                           \text{G} \mathrel{:=} 2\rrparenthesis\big)
                \PC{{\Lmbd{s}{1/2}}} 
                           \Apply\ \big(\Lmbd{s} s\llparenthesis
                           \text{G} \mathrel{:=} 3\rrparenthesis\big)\Big)\\
\text{reveal} &\equiv
    \bigsqcap d \in
        \big(\Lmbd{s}{\{1,2,3\} - \{\text{P}\ s, \text{G}\ s\}}\big).\
        \Apply\ \big(\Lmbd{s} s\llparenthesis
        \text{C} \mathrel{:=} d\rrparenthesis\big) \\
\text{switch} &\equiv
    \bigsqcap d \in
        \big(\Lmbd{s}{\{1,2,3\} - \{\text{C}\ s, \text{G}\ s\}}\big).\
        \Apply\ \big(\Lmbd{s} s\llparenthesis
        \text{G} \mathrel{:=} d\rrparenthesis\big) \\
\\
\text{monty}\ \textit{s}
    &\equiv \text{hide} \Seq \text{guess} \Seq \text{reveal} \Seq
            \If\ \textit{s}\ \Then\ \text{switch}\ \Else\ \Skip
\end{align*}
\caption{\label{f:monty}The Monty Hall game in pGCL}
\end{figure}

Consider \autoref{f:monty}, a model of the Monty Hall
problem\cite{Selvin_75,Hurd_05}.  Briefly, the context is a game show: A
prize is hidden behind one of three doors ($\text{hide}$), of which the
contestant then guesses one ($\text{guess}$).  The host then opens a door
other than the one the contestant has chosen ($\text{reveal}$), showing that
it does not hide the prize.  The contestant now has a choice: to switch to the
unopened door ($\text{switch}$), or to stick to the original ($\Skip$).  Is
the contestant better off switching?

\autoref{f:monty} introduces a new primitive: demonic choice over a set:
\begin{align*}
\bigsqcap x \in \{a,b,\dots\}.\ p\ x &\equiv p\ a \DC p\ b \DC \dots &
\Wp\ \left(\bigsqcap x \in S.\ p\ x\right)\ R\ s
    &= \inf\ \{ \Wp\ (p\ x)\ R\ s \mathop{|} x \in S\ s \}
\end{align*}
The ability to extend the language in this way is a benefit of the shallow
embedding.  We will return to this point in \autoref{s:shallow}.

We first define the victory condition for the game, whence the least
probability of the contestant winning, over all choices by the (demonic) host,
and from a given starting state $s$, is then given by the weakest
precondition:
\begin{gather*}
\text{win}\ g \equiv (\text{G}\ g = \text{P}\ g) \qquad
P_\text{min}(\text{win}) =
    \Wp\ (\text{monty}\ \textit{switch})\ \EB{\text{win}}\ s
\end{gather*}

\paragraph{Proof by unfolding}

When \textit{switch} is false, we solve by explicitly unfolding the rules in
\autoref{f:rules}.  This approach was demonstrated by Hurd et.
al.\cite{Hurd_05} As expected, the contestant has a $1/3$ chance of success:

\medskip

\begin{minipage}{\textwidth}
\begin{isabellebody}
\isacommand{lemma}\isamarkupfalse\ wp\_monty\_noswitch:
"$\lambda s.\ 1/3 \pentails \Wp\ (\text{monty}\ \text{false})\
    \EB{\text{wins}}$"\isanewline
\isadelimproof
\ \ %
\endisadelimproof
\isatagproof
\isacommand{unfolding} monty\_def hide\_def guess\_def
                       reveal\_def switch\_def
\isanewline
\ \ \isacommand{by}(simp add:wp\_eval insert\_Diff\_if)
\endisatagproof
\end{isabellebody}
\end{minipage}


\paragraph{Proof by decomposition}
If \textit{switch} is true the state space grows dramatically, and such a
straightforward proof becomes computationally expensive, although it is still
just possible in this case.  To scale to still larger examples, we must employ
modular reasoning.  Fortunately, weakest preconditions in pGCL admit
composition rules very similar to their counterparts for GCL:
\begin{align*}
&\inferrule*[lab=wp\_strengthen\_post,sep=1em]
    {P \pentails \Wp\ p\ Q \\ Q \pentails R \\ \text{healthy}\ (\Wp p) \\
     \text{sound}\ Q \\ \text{sound}\ R}
    {P \pentails \Wp\ p\ R} \\
&\inferrule*[lab=wp\_Seq,sep=1em]
    {Q \pentails \Wp\ b\ R \\ P \pentails \Wp\ a\ Q \\
     \text{healthy}\ (\Wp\ a) \\ \text{healthy}\ (\Wp\ b) \\
     \text{sound}\ Q \\ \text{sound}\ R}
    {P \pentails \Wp\ (a \Seq b)\ R} \\
\end{align*}
The healthiness and soundness obligations result from the shallow embedding.
The rules are otherwise identical.  To integrate with Isabelle's calculational
reasoner\cite{Bauer_01}, we define probabilistic Hoare triples:
\begin{gather*}
\inferrule*[lab=wp\_validI,sep=1em]
    {P \pentails \Wp\ a\ Q}
    {\valid{P}{a}{Q}} \qquad
\inferrule*[lab=wp\_validD,sep=1em]
    {\valid{P}{a}{Q}}
    {P \pentails \Wp\ a\ Q} \\
\inferrule*[lab=valid\_Seq,sep=1em]
    {\valid{P}{a}{Q} \\ \valid{Q}{b}{R} \\
     \text{healthy}\ (\Wp\ a) \\ \text{healthy}\ (\Wp\ b) \\
     \text{sound}\ Q \\ \text{sound}\ R}
    {\valid{P}{a \Seq b}{R}}
\end{gather*}
Note that \textsc{valid\_Seq} is simply the composition of
\textsc{wp\_compose} with \textsc{wp\_validI} and \textsc{wp\_validD}.  We
need one final rule, peculiar to pGCL, which illustrates the real-valued
nature of expectations:
\begin{gather*}
\inferrule*[lab=wp\_scale,sep=1em]
    {P \pentails \Wp\ a\ Q \\ \text{healthy}\ (\Wp\ a) \\
     \text{sound}\ Q \\ 0 < c}
    {(\lambda s.\ c \times P\ s) \pentails \Wp\ a\ (\lambda s.\ c \times Q\ s)}
\end{gather*}
This follows from the healthiness of the transformer, and allows us to scale
the pre- and post-expectations such that the latter `fits under' some target.
To illustrate, consider the `obvious' specification of hide:
\begin{gather}
\lambda s.\ 1 \pentails \Wp\ \text{hide}\ \EB{\lambda s.\ P\ s \in \{1,2,3\}}
\label{e:wp_hide}
\end{gather}
This states that with probability 1, the prize ends up behind door 1, 2 or 3.
In evaluating our preconditions stepwise, however, we find that the weakest
precondition of the remainder of the program is in fact:
\begin{gather*}
\lambda s.\ \ 2/3 \times \EB{\lambda s.\ P\ s \in \{1,2,3\}}\ s\ \text{.}
\end{gather*}
Applying rule \textsc{wp\_scale} to \autoref{e:wp_hide} we derive a scaled
rule:
\begin{gather*}
\lambda s.\ 2/3 \pentails \Wp\ \text{hide}\ 
    (\lambda s.\ 2/3 \times \EB{\lambda s.\ P\ s \in \{1,2,3\}}\ s)\ \text{.}
\end{gather*}
Using this, we see that the probability of success if the contestant switches
is at least\footnote{In fact it is exactly 2/3, but our object is to
demonstrate an entailment proof.  In more complicated situations, calculating
the exact pre-expectation is impractical.} 2/3:

\medskip

\begin{minipage}{\textwidth}
\begin{isabellebody}
\isacommand{declare}\isamarkupfalse\ valid\_Seq[trans]\isanewline
\isacommand{lemma}\isamarkupfalse\ wp\_monty\_switch\_modular:
"$\lambda s.\ 2/3 \pentails \Wp\ (\text{monty}\ \text{true})\
    \EB{\text{wins}}$"\isanewline
\isadelimproof
\ \ %
\endisadelimproof
\isatagproof
\isacommand{proof}\isamarkupfalse (rule\ wp\_validD)\isanewline
\ \ \ \ \isacommand{note}\ wp\_validI[OF wp\_scale,\ OF wp\_hide,\
                                      simplified]\isanewline
\ \ \ \ \isacommand{also}\ \isacommand{note}\ 
    wp\_validI[OF wp\_guess]\isanewline
\ \ \ \ \isacommand{also}\ \isacommand{note}\ 
    wp\_validI[OF wp\_reveal]\isanewline
\ \ \ \ \isacommand{also}\ \isacommand{note}\ 
    wp\_validI[OF wp\_switch]\isanewline
\ \ \ \ \isacommand{finally}\ \isacommand{show}\ 
    "$\lambda s.\ 2/3 \pentails \Wp\ (\text{monty}\ \text{true})\ 
                                \EB{\text{wins}}$"
    \isanewline
\ \ \ \ \ \ \isacommand{unfolding}\ monty\_def
\isacommand{by}\isamarkupfalse
    (simp\ add:healthy\_intros\ sound\_intros\ monty\_healthy)\isanewline
\ \ \isacommand{qed}
\endisatagproof
\end{isabellebody}
\end{minipage}

\medskip

As mentioned, we take advantage of the calculational reasoning facility of
Isabelle.  The intermediate Hoare triples are automatically derived, by
applying \textsc{valid\_Seq} to the previous relation and the supplied
specification.  Finally, we again discharge all side-conditions using the
simplifier.

\paragraph{Proof by VCG}
Finally, we can pass the component specifications to our verification
condition generator (VCG), which follows a similar strategy to the above,
automatically matching the appropriate rule to the goal.  The VCG leaves an
inequality between the target pre-expectation and that calculated internally
(generally not the weakest).  In this case, the final goal is trivial enough
to be discharged internally.

\medskip

\begin{minipage}{\textwidth}
\begin{isabellebody}
\isacommand{lemmas} scaled\_hide = wp\_scale[OF wp\_hide, simplified]
\isanewline
\isacommand{declare} scaled\_hide[wp] wp\_guess[wp]
                     wp\_reveal[wp] wp\_guess[wp] \isanewline
\isacommand{declare} healthy\_wp\_hide[health] healthy\_wp\_guess[health]
\isanewline
\phantom{\isacommand{declare}} healthy\_wp\_reveal[health]
    healthy\_wp\_switch[health] \isanewline
\isacommand{lemma}\isamarkupfalse\ wp\_monty\_switch\_vcg:
"$\lambda s.\ 2/3 \pentails \Wp\ (\text{monty}\ \text{true})\
    \EB{\text{wins}}$"\isanewline
\isadelimproof
\ \ %
\endisadelimproof
\isatagproof
\ \ \isacommand{unfolding} monty\_def\ \isacommand{by}(simp,pvcg)
\endisatagproof
\end{isabellebody}
\end{minipage}

\subsection{Loops and Recursion}

Recursive programs cannot be evaluated by unfolding.  The treatment of
recursion in pGCL is well developed, and we have incorporated some of this
work, specifically regarding loops.  Our treatment is at an early stage of
development, but we already provide several useful rules, including this,
which is a specialisation of lemma 7.3.1 of McIver \&
Morgan\cite{McIver_M_04}, and gives the correctness condition for standard
post-expectations on loops which terminate\footnote{A strictly weaker
condition than terminating along all paths:  Non-terminating traces with
probability 0 are acceptable.  Consider flipping a coin until it shows heads:
The only non-terminating path (flipping tails forever) has probability 0.}
with probability 1:
\begin{equation}\label{e:looprule}
\inferrule*[lab=wp\_Loop,sep=1em]
    {\text{well\_def}\ (\Wp\ \textit{body}) \\
     \textrm{sub\_distrib}\ (\textbf{do}\ G \rightarrow \textit{body})\\
     (\lambda s.\ \EB{G}\ s \times \EB{I}\ s) \pentails \Wp\ \textit{body}\ \EB{I}}
    {\EB{I} \mathop{\&\&} \Wp\ (\textbf{do}\ G \rightarrow \textit{body})\ 
                          (\lambda s. 1)
     \pentails \Wp\ (\textbf{do}\ G \rightarrow \textit{body})\ 
        (\lambda s. \EB{\overline{G}} \times \EB{I})}
\end{equation}

\section{Implementation and Extensions}\label{s:how}

We now expand on some details of the implementation, and its extensions.

\subsection{Implementing wp and wlp}
\label{s:wp_impl}

\begin{figure}
\begin{align*}
a \Seq b &\equiv \Lmbd{ab}{(a\ ab) \circ (b\ ab)}
&
\Abort &\equiv \Lmbd{ab\,P}{\If\ ab\ \Then\ \Lmbd{s}{0}\
                               \Else\ \Lmbd{s}{\textrm{bound\_of}\ P}}
\\
\Embed\ f &\equiv \Lmbd{ab}{f}
&
\Mu{x}{\textrm{prog}\ x} &\equiv
  \Lmbd{ab}{\If\ ab\
  \begin{array}[t]{l}
    \Then\ \textrm{lfp\_trans}\ (\Lmbd{t}{\textrm{prog}\ (\Embed\ t)\ ab})\\
    \Else\ \textrm{gfp\_trans}\ (\Lmbd{t}{\textrm{prog}\ (\Embed\ t)\ ab})
   \end{array}}
\\
\Wp\ a &\equiv a\ \text{True}
&
\Wlp\ a &\equiv a\ \text{False}
\end{align*}
\caption{\label{f:wp_impl} The underlying definitions of selected primitives}
\end{figure}

\autoref{f:wp_impl} details the implementation of several primitives, together
with the definitions of wp and wlp.  Programs are represented as their
associated transformer, parameterised by the treatment of $\Abort$ (the
parameter $ab$), giving either strict or liberal semantics.  Only $\Abort$ and
$\mu$ change their behaviour between wp and wlp: The former gives either
failure ($\lambda P\,s.\ 0$) or success ($\lambda P\,s.\ \text{bound\_of}\
P$), whereas the latter is the least or the greatest fixed point,
respectively.  All others, as for ($\Seq$), and simply pass $ab$ inward.

\subsection{Consequences of a Shallow Embedding}\label{s:shallow}

The very shallow embedding used has two important consequences, the first of
which is negative. The healthiness of transformers, and soundness of
expectations, must be explicitly carried as assumptions.  A deeper
embedding\cite{Hurd_05} might restrict to the type of healthy transformers,
in which case these would be satisfied by the type axioms.

We avoid such an embedding to reuse as much of the mechanisation within
Isabelle/HOL as possible.  Reasoning within a fresh type requires lifting (and
modifying) all necessary rules.  The burden of discharging our side-conditions
is, moreover, not high.  For any primitively constructed program, healthiness
follows by invoking the simplifier with the appropriate lemmas.

The positive consequence of an extremely shallow embedding is the ease with
which it can be extended.  We have already seen an example: the definition of
demonic choice from a set (following a standard
abbreviation\cite{McIver_M_04}).  To do so, one need only supply
weakest-precondition (and weakest-liberal-precondition) rules, rules to infer
healthiness and (optionally) rules for proof decomposition.

Interestingly, it is not necessary to show that the new primitive is sound,
that is, produces a healthy transformer for all inputs.  It is merely
necessary that the supplied rules show healthiness for just those cases in
which it is actually used.  Set demonic choice is just such a partially sound
primitive: Healthiness does not generally hold for infinite sets.  Applied
here to finite sets, the given rules establish healthiness.

\begin{figure}
\begin{align*}
&\textbf{type\_synonym}\ (\sigma,\alpha)\ \text{nondet\_monad} =
    \sigma \Rightarrow (\alpha \times \sigma)\ \text{set} \times \text{bool}
    \\
&\lbrace P\rbrace f \lbrace Q\rbrace \equiv
    \forall s.\ P\ s \longrightarrow
        (\forall(r,s') \in \text{fst}\ (f\ s).\ Q\ r\ s') \\
&\text{no\_fail}\ P\ m \equiv
    \forall s.\ P\ s \longrightarrow \lnot (\text{snd}\ (m\ s))\\
\\
&\textrm{Exec} :: (\sigma,\alpha)\ \text{nondet\_monad} \Rightarrow
                  \sigma\ \text{prog}\\
&\textrm{Exec}\ M \equiv \lambda ab\ R\ s.\\
&\quad\textrm{let}\ (SA,f) = M\ s\ \textrm{in} &\textit{Run the monad} \\
&\quad\quad\textrm{if}\ f\ \textrm{then}\ \Abort\ ab\ R\ s
    &\textit{Fail is Abort} \\
&\quad\quad\phantom{\textrm{if}\ f}\ \textrm{else}\ 
    \textrm{if}\ SA = \{\}\ \textrm{then}\ (\text{bound\_of}\ R)
        &\textit{Stuck is Success} \\
&\quad\quad\phantom{\textrm{if}\ f\ \textrm{else}\ 
    \textrm{if}\ SA = \{\}}\ 
    \textrm{else}\ \textrm{let}\ S = \textrm{snd} \mathop{\char"12} SA\ 
        \textrm{in} &\textit{Ignore result} \\
&\quad\quad\phantom{\textrm{if}\ f\ \textrm{else}\ 
    \textrm{if}\ SA = \{\}\ \textrm{else}}\ 
    \textrm{glb}\ (R \mathop{\char"12} S)
        &\textit{Infimum over states}
\end{align*}
\caption{\label{f:nd_orig}The L4.verified non-deterministic monad in Isabelle}
\end{figure}

As a further example, we embed the non-deterministic monad at the heart of the
L4.verified proof.  The existing definition, \autoref{f:nd_orig}, is as a
function from states ($\sigma$) to a set of result ($\alpha$), state pairs.
The extra result is the failure flag, used to explicitly signal failure.  This
was added to ensure that termination is preserved under refinement, as
described elsewhere\cite{Cock_KS_08}.  We embed as follows: Stuck (no
successor states) is success, for compatibility with our
infimum-over-alternatives interpretation;  Explicit failure is $\Abort$, which
in turn is either success or failure under $\Wlp$ or $\Wp$, respectively.  We
lift results as follows:
\begin{align*}
&\inferrule*[lab=wp\_Exec,sep=1em]
    {\lbrace P \rbrace\ \textit{prog}\ \lbrace \lambda r\ s.\ Q\ s\rbrace \\
     \textrm{no\_fail}\ P\ \textit{prog} \\
     \exists s.\ P\ s}
    {\EB{P} \entails \Wp\ \textit{prog}\ \EB{Q}} &
&\inferrule*[lab=wlp\_Exec,sep=1em]
    {\lbrace P \rbrace\ \textit{prog}\ \lbrace \lambda r\ s.\ Q\ s\rbrace \\
     \exists s.\ P\ s}
    {\EB{P} \entails \Wlp\ \textit{prog}\ \EB{Q}} \\
\end{align*}

\subsection{Fixed Points and the Lattice Structure of Expectations and
            Transformers}\label{s:lattice}

Handling recursion means reasoning about fixed points.  In this case, we need
both least and greatest, on expectations and on transformers.  Due to the
shallow embedding, we cannot appeal to the existing fixed point results, which
are phrased on a complete lattice.  Neither the underlying type for
expectations ($S \rightarrow \mathbb{R}$) or for transformers ($(S \rightarrow
\mathbb{R}) \rightarrow S \rightarrow \mathbb{R}$) can be so instantiated, due
to the lack of both top and bottom elements.  The solution in each case is
different.

Sound expectations have an obvious bottom element, $\lambda s.\ 0$, but there
is no universal upper bound.  We only require that there exists a bound for
any given expectation.  There need not exist any bound on an arbitrary set of
sound expectations.  For example, with $S = \mathbb{N}$, consider the set
$$
\{ (\lambda s.\ \If\ s=n\ \Then\ n\ \Else\ 0) : n \in \mathbb{N}\}\ \text{.}
$$
Each expectation is bounded (by $n$) and non-negative, yet the least upper
bound, $\lambda s.\ s$, is unbounded.

We need a surrogate for the top element.  To illustrate, take our definition
for greatest fixed point:
\begin{align*}
\text{gfp\_in}\ f\ S &\equiv \If\ \exists x\in S.\ x \le f\ x\ \Then\ \text{lub}\
    \{x\in S.\ x \le f\ x\}\ \Else\ \text{lowerbound}\ S
\end{align*}

The lower bound is easy ($\lambda s.\ 0$), but in order to find the {\em
least} upper bound, we first need {\em some} upper bound.  In a complete
lattice, this is the top element.  Instead, we appeal to feasibility:
$$
\text{bound\_of}\ ((\Mu{x}{f\ x})\ P) \le \text{bound\_of}\ P\ \text{,}
\quad\text{and thus}\quad
(\Mu{x}{f\ x})\ P \le \lambda s.\ \text{bound\_of}\ P\ \text{.}
$$
It is therefore sufficient to consider fixed points that are {\em weakly
bounded} by $P$:
\begin{gather*}
\text{weakly\_bounded\_by}\ P \equiv
    \{Q.\ \text{sound}\ Q \land \text{bound\_of}\ Q \le \text{bound\_of}\ P\}
\end{gather*}
Finally, we establish the standard fixed-point results parameterised by $P$.
For example:
$$
\inferrule*[lab=gfp\_in\_unfold,sep=1em]
    {\text{healthy}\ t}
    {\text{gfp\_in}\ t\ (\text{weakly\_bounded\_by}\ P)
        = t\ (\text{gfp\_in}\ t\ (\text{weakly\_bounded\_by}\ P))}
$$

The case of transformers is simpler, again due to feasibility:
$$
t\ P\ s \le \text{bound\_of}\ P\quad\text{and thus}\quad
t \le \lambda P\ s.\ \text{bound\_of}\ P
$$
Thus we have a top element, and establish a complete lattice by means of a
quotient:
\begin{gather*}
\text{le\_trans}\ t\ u \equiv
    \forall P.\ \text{sound}\ P \rightarrow t\ P \le u\ P \qquad
\text{equiv\_trans}\ t\ u \equiv
    \text{le\_trans}\ t\ u \land \text{le\_trans}\ u\ t \\
\text{htrans\_rel}\ t\ u \equiv
    \text{healthy}\ t \land \text{healthy}\ u \land
    \text{equiv\_trans}\ t\ u \\
\textbf{quotient\_type}\ \sigma\ \text{trans} =
    (\sigma \rightarrow \mathbb{R}) \rightarrow
     \sigma \rightarrow \mathbb{R} \mathop{/}
     \text{partial} : \text{htrans\_rel}
\end{gather*}
Using the induced homomorphism, we draw back the standard results:
$$
\inferrule*[lab=gfp\_trans\_unfold,sep=1em]
    {\wedge t.\ \text{healthy}\ t \entails \text{healthy}\ (T\ t)\\
    \wedge t\ u.\ [\text{healthy}\ t; \text{healthy}\ u;
                   \text{le\_trans}\ t\ u] \entails
                  \text{le\_trans}\ (T\ t)\ (T\ u)}
    {\text{equiv\_trans}\ (\text{gfp\_trans}\ T)\ (T\ (\text{gfp\_trans}\ T))}
$$

\subsection{The Verification Condition Generator}
\label{s:vcg}

The VCG tactic, \texttt{pvcg}, is simple but nonetheless capable of handling
\autoref{f:monty}.  It alternates applying an entailment rule, and attempting
to discharge side-goals using internal and user-supplied rules.

The user supplies specifications as proved entailment rules, tagged with
\texttt{[wp]}, and healthiness rules, tagged with \texttt{[health]}.  Internal
rules are tagged \texttt{[wp\_core]}.  The VCG selects rules as follows:
\begin{enumerate}
\item User-supplied rules, as written.
\item User-supplied rules, with a strengthened postcondition:
\texttt{wp\_strengthen\_post[OF rule]}.  This leaves a side-goal that the
postcondition of the supplied rule entails the strengthened version.
\item Internal rules.
\end{enumerate}
A user-supplied rule will override an internal rule, and may refer directly to
a compound structure e.g.~$P \pentails \Wp\ (a \Seq b)\ Q$.  The given rule
will be used rather than unfolding the composition.  If no user rule is found,
the VCG will proceed using its internal rules, calculating the exact weakest
precondition by unfolding.

\section{Related Work}
\label{s:related}

The mechanisations of many existing programming logic formalisations have been
presented, in Isabelle and in other theorem
provers\cite{Nipkow_02,Mossakowski_10,Cock_KS_08,Harrison_05}.  Relative to
these, the novelty of this work is in the treatment of probabilistic programs
and properties, through pGCL.

A previous formalisation of pGCL, in the HOL4 theorem prover, was presented by
Hurd et.~al.\cite{Hurd_05}, whence we have adapted our worked example (Monty
Hall).  Celiku \& McIver\cite{Celiku_04}, working from this same
formalisation, developed an approach to analysing the execution time of
probabilistic algorithms, presenting a mechanised proof of the
self-stabilisation time for Herman's ring.  We have presented a novel approach
to formalising the same underlying logic.  By prioritising the reuse of
existing theories in Isabelle/HOL\cite{Nipkow_PW:Isabelle}, in particular by
defining expectations using the standard real type by means of our quotient
and pullback technique, we achieve excellent automation and integration with
existing work, notably L4.verified\cite{Klein_EHACDEEKNSTW_09}.

\section{Applications \& Ongoing Work}\label{s:future}
\label{s:ongoing}

This formalisation was developed to assist in the verification of
high-assurance component systems.  Our principal application is the
verification of probabilistic security properties, of the sort established in
\autoref{s:why}.  Having established that a guessing-attack based measure is
refinement-sound, we plan to formally establish bounds on vulnerability due to
guessing attacks involving side-channel leakage.

The fact that we can embed the nondeterministic monad used in the L4.verified
proof is significant, as it means that we can appeal directly to that
project's top-level theorem in establishing a probabilistic result on seL4
(the microkernel in question).  An attractive target is a randomised
scheduler: In seL4, as in many kernels, the scheduler is a small component,
called at the top level after all other state-modifying code has executed.  We
could thus appeal to the existing, unmodified, proof to establish the
soundness of the transformation and then, lifting the result via the above
construction, establish a probabilistic result on the system including the
scheduler.  As randomised scheduling is a common approach to side-channel
mitigation, this is a promising approach to achieving our stated goal of
formally establishing side-channel bounds in real systems.

\clearpage
\phantomsection
\addcontentsline{toc}{section}{\refname}
\bibliographystyle{eptcs}

\bibliography{references}

\providecommand{\noopsort}[1]{} \providecommand{\url}{\error{The bib files now
  require `url' package!}}
\begin{thebibliography}{10}
\providecommand{\bibitemdeclare}[2]{}
\providecommand{\surnamestart}{}
\providecommand{\surnameend}{}
\providecommand{\urlprefix}{Available at }
\providecommand{\url}[1]{\texttt{#1}}
\providecommand{\href}[2]{\texttt{#2}}
\providecommand{\urlalt}[2]{\href{#1}{#2}}
\providecommand{\doi}[1]{doi:\urlalt{http://dx.doi.org/#1}{#1}}
\providecommand{\bibinfo}[2]{#2}

\bibitemdeclare{inproceedings}{Bauer_01}
\bibitem{Bauer_01}
\bibinfo{author}{Gertrud \surnamestart Bauer\surnameend} \&
  \bibinfo{author}{Markus \surnamestart Wenzel\surnameend}
  (\bibinfo{year}{2001}): \emph{\bibinfo{title}{Calculational Reasoning
  Revisited -- An Isabelle/Isar experience}}.
\newblock In: {\sl \bibinfo{booktitle}{14th TPHOLs}},
  \bibinfo{publisher}{Springer-Verlag}, pp. \bibinfo{pages}{75--90},
  \doi{10.1007/3-540-44755-5\_7}.

\bibitemdeclare{article}{Celiku_04}
\bibitem{Celiku_04}
\bibinfo{author}{Orieta \surnamestart Celiku\surnameend} \&
  \bibinfo{author}{Annabelle \surnamestart McIver\surnameend}
  (\bibinfo{year}{2004}): \emph{\bibinfo{title}{Cost-based analysis of
  probabilistic programs mechanised in {HOL}}}.
\newblock {\sl \bibinfo{journal}{Nordic J. of Computing}}
  \bibinfo{volume}{11}(\bibinfo{number}{2}), pp. \bibinfo{pages}{102--128}.
\newblock \urlprefix\url{http://www.cse.unsw.edu.au/~carrollm/probs/Papers/
  Celiku-04.pdf}.

\bibitemdeclare{inproceedings}{Cock_11}
\bibitem{Cock_11}
\bibinfo{author}{David \surnamestart Cock\surnameend} (\bibinfo{year}{2011}):
  \emph{\bibinfo{title}{Exploitation as an inference problem}}.
\newblock In: {\sl \bibinfo{booktitle}{4th AISEC}}, \bibinfo{address}{Chicago,
  IL, USA}, pp. \bibinfo{pages}{105--106}, \doi{10.1145/2046684.2046702}.

\bibitemdeclare{inproceedings}{Cock_KS_08}
\bibitem{Cock_KS_08}
\bibinfo{author}{David \surnamestart Cock\surnameend}, \bibinfo{author}{Gerwin
  \surnamestart Klein\surnameend} \& \bibinfo{author}{Thomas \surnamestart
  Sewell\surnameend} (\bibinfo{year}{2008}): \emph{\bibinfo{title}{Secure
  Microkernels, State Monads and Scalable Refinement}}.
\newblock In \bibinfo{editor}{Otmane~Ait \surnamestart Mohamed\surnameend},
  \bibinfo{editor}{C\'{e}sar \surnamestart Mu{\~{n}}oz\surnameend} \&
  \bibinfo{editor}{Sofi\`{e}ne \surnamestart Tahar\surnameend}, editors: {\sl
  \bibinfo{booktitle}{21st TPHOLs}}, {\sl \bibinfo{series}{LNCS}}
  \bibinfo{volume}{5170}, \bibinfo{publisher}{Springer-Verlag},
  \bibinfo{address}{Montreal, Canada}, pp. \bibinfo{pages}{167--182},
  \doi{10.1007/978-3-540-71067-7\_16}.

\bibitemdeclare{article}{Dijkstra_75}
\bibitem{Dijkstra_75}
\bibinfo{author}{Edsger~W. \surnamestart Dijkstra\surnameend}
  (\bibinfo{year}{1975}): \emph{\bibinfo{title}{Guarded commands,
  nondeterminacy and formal derivation of programs}}.
\newblock {\sl \bibinfo{journal}{CACM}}
  \bibinfo{volume}{18}(\bibinfo{number}{8}), pp. \bibinfo{pages}{453--457},
  \doi{10.1145/360933.360975}.

\bibitemdeclare{inproceedings}{Goguen_Meseguer_82}
\bibitem{Goguen_Meseguer_82}
\bibinfo{author}{Joseph \surnamestart Goguen\surnameend} \&
  \bibinfo{author}{Jos{\'e} \surnamestart Meseguer\surnameend}
  (\bibinfo{year}{1982}): \emph{\bibinfo{title}{Security Policies and Security
  Models}}.
\newblock In: {\sl \bibinfo{booktitle}{IEEE Symp. Security \& Privacy}},
  \bibinfo{publisher}{Comp. Soc.}, \bibinfo{address}{Oakland, California, USA},
  pp. \bibinfo{pages}{11--20}.

\bibitemdeclare{article}{Harrison_05}
\bibitem{Harrison_05}
\bibinfo{author}{William~L. \surnamestart Harrison\surnameend} \&
  \bibinfo{author}{Richard~B. \surnamestart Kieburtz\surnameend}
  (\bibinfo{year}{2005}): \emph{\bibinfo{title}{The logic of demand in
  Haskell}}.
\newblock {\sl \bibinfo{journal}{J. Functional Progr.}}
  \bibinfo{volume}{15}(\bibinfo{number}{6}), pp. \bibinfo{pages}{837--891},
  \doi{10.1017/S0956796805005666}.

\bibitemdeclare{article}{Hurd_05}
\bibitem{Hurd_05}
\bibinfo{author}{Joe \surnamestart Hurd\surnameend}, \bibinfo{author}{Annabelle
  \surnamestart McIver\surnameend} \& \bibinfo{author}{Carroll \surnamestart
  Morgan\surnameend} (\bibinfo{year}{2005}):
  \emph{\bibinfo{title}{Probabilistic guarded commands mechanized in HOL}}.
\newblock {\sl \bibinfo{journal}{Theoretical Computer Science}}
  \bibinfo{volume}{346}(\bibinfo{number}{1}), pp. \bibinfo{pages}{96 -- 112},
  \doi{10.1016/j.tcs.2005.08.005}.
\newblock \urlprefix\url{http://www.sciencedirect.com/science/article/pii/
  S0304397505004767}.

\bibitemdeclare{inproceedings}{Klein_EHACDEEKNSTW_09}
\bibitem{Klein_EHACDEEKNSTW_09}
\bibinfo{author}{Gerwin \surnamestart Klein\surnameend}, \bibinfo{author}{Kevin
  \surnamestart Elphinstone\surnameend}, \bibinfo{author}{Gernot \surnamestart
  Heiser\surnameend}, \bibinfo{author}{June \surnamestart
  Andronick\surnameend}, \bibinfo{author}{David \surnamestart Cock\surnameend},
  \bibinfo{author}{Philip \surnamestart Derrin\surnameend},
  \bibinfo{author}{Dhammika \surnamestart Elkaduwe\surnameend},
  \bibinfo{author}{Kai \surnamestart Engelhardt\surnameend},
  \bibinfo{author}{Rafal \surnamestart Kolanski\surnameend},
  \bibinfo{author}{Michael \surnamestart Norrish\surnameend},
  \bibinfo{author}{Thomas \surnamestart Sewell\surnameend},
  \bibinfo{author}{Harvey \surnamestart Tuch\surnameend} \&
  \bibinfo{author}{Simon \surnamestart Winwood\surnameend}
  (\bibinfo{year}{2009}): \emph{\bibinfo{title}{{seL4}: Formal Verification of
  an {OS} Kernel}}.
\newblock In: {\sl \bibinfo{booktitle}{22nd SOSP}}, \bibinfo{publisher}{ACM},
  \bibinfo{address}{Big Sky, MT, USA}, pp. \bibinfo{pages}{207--220},
  \doi{10.1145/1629575.1629596}.

\bibitemdeclare{article}{McIver_01}
\bibitem{McIver_01}
\bibinfo{author}{Annabelle \surnamestart McIver\surnameend} \&
  \bibinfo{author}{Carroll \surnamestart Morgan\surnameend}
  (\bibinfo{year}{2001}): \emph{\bibinfo{title}{Partial correctness for
  probabilistic demonic programs}}.
\newblock {\sl \bibinfo{journal}{Theoretical Comp. Sci.}}
  \bibinfo{volume}{266}(\bibinfo{number}{1–2}), pp. \bibinfo{pages}{513 --
  541}, \doi{10.1016/S0304-3975(00)00208-5}.
\newblock \urlprefix\url{http://www.sciencedirect.com/science/article/pii/
  S0304397500002085}.

\bibitemdeclare{book}{McIver_M_04}
\bibitem{McIver_M_04}
\bibinfo{author}{Annabelle \surnamestart McIver\surnameend} \&
  \bibinfo{author}{Carroll \surnamestart Morgan\surnameend}
  (\bibinfo{year}{2004}): \emph{\bibinfo{title}{Abstraction, Refinement and
  Proof for Probabilistic Systems}}.
\newblock \bibinfo{publisher}{Springer}.

\bibitemdeclare{article}{Mossakowski_10}
\bibitem{Mossakowski_10}
\bibinfo{author}{Till \surnamestart Mossakowski\surnameend},
  \bibinfo{author}{Lutz \surnamestart Schr\"oder\surnameend} \&
  \bibinfo{author}{Sergey \surnamestart Goncharov\surnameend}
  (\bibinfo{year}{2010}): \emph{\bibinfo{title}{A generic complete dynamic
  logic for reasoning about purity and effects}}.
\newblock {\sl \bibinfo{journal}{Formal Aspects Comput.}}
  \bibinfo{volume}{22}(\bibinfo{number}{3-4}), pp. \bibinfo{pages}{363--384},
  \doi{10.1007/s00165-010-0153-4}.

\bibitemdeclare{inproceedings}{Nipkow_02}
\bibitem{Nipkow_02}
\bibinfo{author}{Tobias \surnamestart Nipkow\surnameend}
  (\bibinfo{year}{2002}): \emph{\bibinfo{title}{Hoare Logics in
  {Isabelle/HOL}}}.
\newblock In \bibinfo{editor}{H.~\surnamestart Schwichtenberg\surnameend} \&
  \bibinfo{editor}{R.~\surnamestart Steinbr\"uggen\surnameend}, editors: {\sl
  \bibinfo{booktitle}{Proof and System-Reliability}},
  \bibinfo{publisher}{Kluwer}, pp. \bibinfo{pages}{341--367}.

\bibitemdeclare{book}{Nipkow_PW:Isabelle}
\bibitem{Nipkow_PW:Isabelle}
\bibinfo{author}{Tobias \surnamestart Nipkow\surnameend},
  \bibinfo{author}{Lawrence \surnamestart Paulson\surnameend} \&
  \bibinfo{author}{Markus \surnamestart Wenzel\surnameend}
  (\bibinfo{year}{2002}): \emph{\bibinfo{title}{{Isabelle/HOL} --- A Proof
  Assistant for Higher-Order Logic}}.
\newblock {\sl \bibinfo{series}{LNCS}} \bibinfo{volume}{2283},
  \bibinfo{publisher}{Springer-Verlag}.

\bibitemdeclare{article}{Selvin_75}
\bibitem{Selvin_75}
\bibinfo{author}{Steve \surnamestart Selvin\surnameend} (\bibinfo{year}{1975}):
  \emph{\bibinfo{title}{A problem in probability (letter to the editor)}}.
\newblock {\sl \bibinfo{journal}{American Statistician}}
  \bibinfo{volume}{29}(\bibinfo{number}{1}), p.~\bibinfo{pages}{67}.

\end{thebibliography}

\end{document}